\documentclass[journal]{IEEEtran}
\usepackage{graphicx}
\usepackage{amssymb}
\usepackage{amsfonts}
\usepackage{amsmath}
\usepackage{epsfig}
\usepackage{color}
\usepackage{fancybox}
\usepackage{textcomp}
\usepackage{multirow}
\usepackage{setspace}
\usepackage{psfrag}
\usepackage{booktabs}
\usepackage{float}
\usepackage{array}

\usepackage{color}

\begin{document}
	%

	\title{Lightweight Convolutional Neural Networks for CSI Feedback in Massive MIMO}
	%
	%
	%
	\author{Zheng~Cao,~Wan-Ting~Shih,~Jiajia~Guo,~Chao-Kai~Wen,~\textit{Member, IEEE},~Shi~Jin,~\textit{Senior Member, IEEE}
			    \thanks{Z. Cao, J. Guo and S. Jin are with the National Mobile Communications Research Laboratory,
			    Southeast University, Nanjing, 210096, P. R. China (e-mail: 707647403@qq.com, \{jiajiaguo, jinshi\}@seu.edu.cn).}
				\thanks{W.-T. Shih and C.-K. Wen are with the Institute of Communications Engineering, National Sun Yat-sen University,
				Kaohsiung 80424, Taiwan (e-mail: sydney2317076@gmail.com, chaokai.wen@mail.nsysu.edu.tw).}
	}
	\vspace{-3cm}
	\maketitle
	\begin{abstract}
		In frequency division duplex mode of massive multiple-input multiple-output systems, the downlink channel state information (CSI) must be sent to the base station (BS) through a feedback link. However, transmitting CSI to the BS is costly due to the bandwidth limitation of the feedback link. Deep learning (DL) has recently achieved remarkable success in CSI feedback. Realizing high-performance and low-complexity CSI feedback is a challenge in DL based communication. We develop a DL based CSI feedback network in this study to complete the feedback of CSI effectively. However, this network cannot be effectively applied to the mobile terminal because of the excessive numbers of parameters. Therefore, we further propose a new lightweight CSI feedback network based on the developed network. Simulation results show that the proposed CSI network exhibits better reconstruction performance than that of other CsiNet-related works. Moreover, the lightweight network maintains a few parameters and parameter complexity while ensuring satisfactory reconstruction performance. These findings suggest the feasibility and potential of the proposed techniques.
	\end{abstract}
	\setlength{\parskip}{-1em}
	%
	\setlength{\parskip}{0em}
	\IEEEpeerreviewmaketitle
\vspace{-0.3cm}
	\section{Introduction}
	Massive multiple-input multiple-output (MIMO) systems are widely regarded as the main technology of 5G wireless communication systems \cite{boccardi2014five}. These systems refer to a communication system that uses hundreds of antennas at the same time-frequency resource to serve tens of user equipment (UE) simultaneously. In frequency division duplex (FDD) mode, a massive MIMO system must feedback the downlink channel state information (CSI) to the base station (BS) to realize its potential gain. However, transmitting CSI to the BS is costly due to the bandwidth limitation.


	In recent years, deep learning (DL) technology has been widely applied in the field of wireless communication \cite{wang2017deep, chen2020jiyu}. The authors in \cite{wen2018deep} used DL technology to build a new type of CSI sensing and recovery neural network called CsiNet. This network can learn how to use the channel architecture effectively for the conversion from CSI to codeword and vice versa. The reconstruction performance of CsiNet is superior to that of traditional compressed sensing methods. Subsequent related studies expanded the original scope of the network. \cite{wang2018deep} considered the time correlation to improve the reconstruction performance of the network. \cite{liu2019exploiting} utilized the correlation between the up-link and downlink to reduce the CSI feedback payload remarkably. \cite{li2020spatio} used a deep recurrent neural network to learn temporal correlation and adopted deep separable convolution to shrink the model, thereby improving the reconstruction performance. \cite{jang2019deep} minimized the effect of feedback transmission errors and delays. \cite{guo2020convolutional} proposed a novel quantization CSI feedback network that considers non-uniform $\mu$-law quantization to obtain uniformly distributed quantization symbols. \cite{yang2019deep} introduced a quantization-based entropy coding with an architecture comprising convolutional layers, followed by quantization and entropy coding blocks. \cite{guo2020dl} proposed a DL-based multi-user collaborative feedback architecture, which is a distributed feedback. The encoders of two nearby UEs cooperate to extract the information of different channel paths under this framework, thereby reducing the overhead of repeated feedback of shared information. \cite{liu2020adversarial} considered the security issues of DL-based CSI feedback. \cite{ye2020deep} investigated interference and non-linear effects in practical scenarios.

	Previous studies suggested that the architecture of the CSI feedback network is roughly divided into the feedforward neural network and the convolutional neural network (CNN). The latter exhibits more satisfactory performance than the former, but with extremely high computational complexity. We use a convolutional autoencoder in the present study to develop the network structure and complete the CSI feedback to preserve the spatial information of 2D signals effectively. The implementation of the CSI feedback network on the mobile terminal is challenging because this network, which is based on the CNN architecture, has excessive numbers of parameters and computational complexity. Thus, we adopt a lightweight structure based on the developed network to build a new one to reduce the complexity and number of parameters. The contributions of this study are summarized as follows.

    We propose a DL-based CSI feedback structure called ConvCsiNet for FDD MIMO systems. This structure uses convolutional layers to extract features, while the mean-pooling and upsampling layers are used to compress and expand the matrix dimension multiples. Considering the application of mobile devices in the field of modern communications, we also propose a DL-based CSI feedback lightweight structure called ShuffleCsiNet. This structure can acquire accurate downlink CSI while consuming low memory space and core computing power. The experimental results show that ConvCsiNet is superior to CsiNet and other CsiNet-related works considering reconstruction performance. Moreover, ConvCsiNet improves the performance of the network at the cost of complexity. Compared with ConvCsiNet, the parameter number and algorithm complexity of ShuffleCsiNet are low, while the reconstruction performance is only slightly degraded.

    \begin{figure*}
		\centering
		\includegraphics[width=7in,height=1.75in]{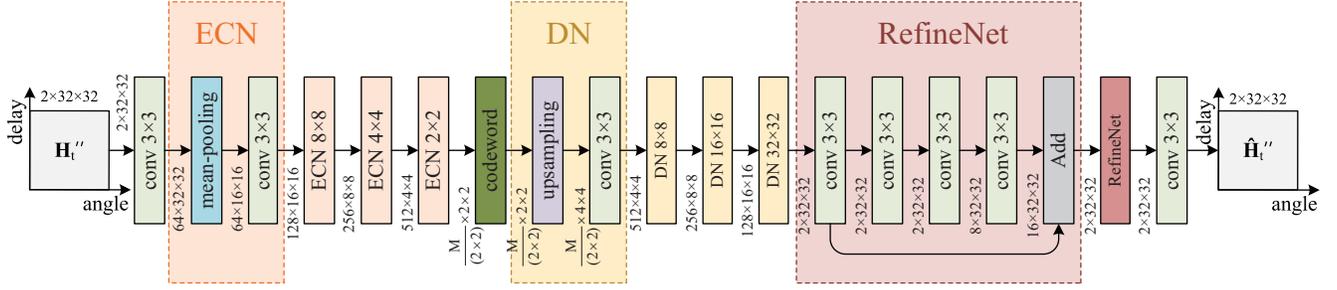}
		\vspace{-0.45cm}
		\caption{Architecture of ConvCsiNet: An encoder structure constructed with convolutional layers and encoded convolution network (ECN) units, which include mean-pooling and convolutional layers; a decoder structure with convolutional layers, two series-connected RefineNet units, and DN units, which include upsampling and convolutional layers ({\bf conv}: convolutional layer)}
		\vspace{-0.5cm}
		\label{ConvCsiNet}
	\end{figure*}

\vspace{-0.3cm}
	\section{System model}
	\label{system_model}
	This study considers a simple massive single-cell FDD downlink MIMO system with a BS and a UE. The BS is configured with a uniform linear antenna array, the number of transmitting antennas is $N_ {t} \gg 1$, and the UE uses a single receive antenna. An OFDM system, which transmits information on $N_ {c}$ orthogonal subcarriers, is also considered. In a time-varying channel, the signal on the $n$ ($n=1,2, \ldots, N_{c}$) subcarrier received by the UE can be expressed as follows:



	\vspace{-0.1cm}
	\begin{equation}
		\begin{aligned}
			{y_{n}}={{\bf h}_{n}^{T}{\bf v}_{n}x_{n}+z_{n}},
			\label{mimo-ofdm}
		\end{aligned}
	\end{equation}
	where ${\bf h}_{n}\in \mathbb{C}^{N_{t}\times 1}$, ${\bf v}_{n} \in \mathbb{C}^{N_{t}\times 1}$, $x_{n}\in \mathbb{C}$, and $z_{n}\in \mathbb{C}$ respectively denote the instantaneous channel vector in the frequency domain, precoding vector, data symbol transmitted in the downlink, and additive Gaussian white noise. The channel vector on all subcarriers is the instantaneous downlink CSI matrix necessary for the UE to facilitate feedback. At this time, the CSI matrix can be denoted as ${\bf H} = [{\bf h}_{1}, \ldots, {\bf h}_{N_{c}}]^{T} \in \mathbb{C}^{N_{c} \times N_{t}}$. After receiving the CSI matrix ${\bf H}$ fed back by the UE, the BS can design the ${\bf v}_{n}$ precoding vector to reduce the interference among users. In actual situations, the UE must continuously estimate and feedback the instantaneous CSI to the BS to enable tracking of changes in the time-varying channel and adjust the corresponding precoding matrix. A total of $N_{t} N_{c}$ complex numbers must be transmitted for a complete CSI matrix due to a large number of antennas. These data occupy a substantial amount of feedback resources, which is undesirable for massive MIMO FDD systems in practical situations. Using a 2D discrete Fourier transform (2D-DFT), the CSI matrix ${\bf H}$ can be converted into an angular-delay domain matrix denoted as \cite{wen2018deep}

	\vspace{-0.1cm}
	\begin{equation}
		\begin{aligned}
			{\bf H}'={\bf F}_{\sf d}{\bf H}{\bf F}_{\sf a},
			\label{CSI-2DDFT}
		\end{aligned}
	\end{equation}
    where ${\bf F}_{\sf d} \in \mathbb{C}^{N_{c} \times N_{c}}$ and ${\bf F}_{\sf a} \in \mathbb{C}^{N_{t} \times N_{t}}$ are two DFT matrices. As proven in \cite{wen2018deep}, the CSI matrix ${\bf H}'$ is sparse in the angular-delay domain when performing DFT on the spatial domain channel vectors (i.e., row vectors of ${\bf H}$) if the number of transmit antennas $N_{t} \to +\infty $ is excessively large \cite{wen2014channel}. Only the first $N_{t}'$ row contains non-zero values for the angular-delay domain channel matrix ${\bf H}'$. Therefore, we only keep the first $N_{t}'$ rows of ${\bf H}'$ and obtain ${\bf H}'' \in \mathbb{C}^{N_{c}' \times N_{t}}$. The total number of parameters for the feedback is then reduced to $N=N_{c}'N_{t}$.

	We mainly consider the autoencoder network for the downlink CSI feedback in this study. The channel matrix ${\bf H}''$ is transmitted to the autoencoder network, and the encoder compresses this matrix into codeword ${\bf s}$ according to a given compression ratio. After ${\bf s}$ is fed back to the BS, the decoder reconstructs ${\bf s}$ to ${\bf H}''$.



\vspace{-0.3cm}
	\section{Two proposed DL-based CSI networks}
	\label{Two proposed DL-based CSI networks}
	We develop the DL-based CSI feedback network in this section using a convolutional autoencoder. On this basis, we then propose a new lightweight structured CSI feedback network. The specific model structure and parameter details are presented in the following subsections.

	\subsection{CSI feedback neural network architecture based on convolutional autoencoder}
	\setlength{\parskip}{0.1em}

	\begin{figure*}
		\centering
		\includegraphics[width=3.98in,height=2.227in]{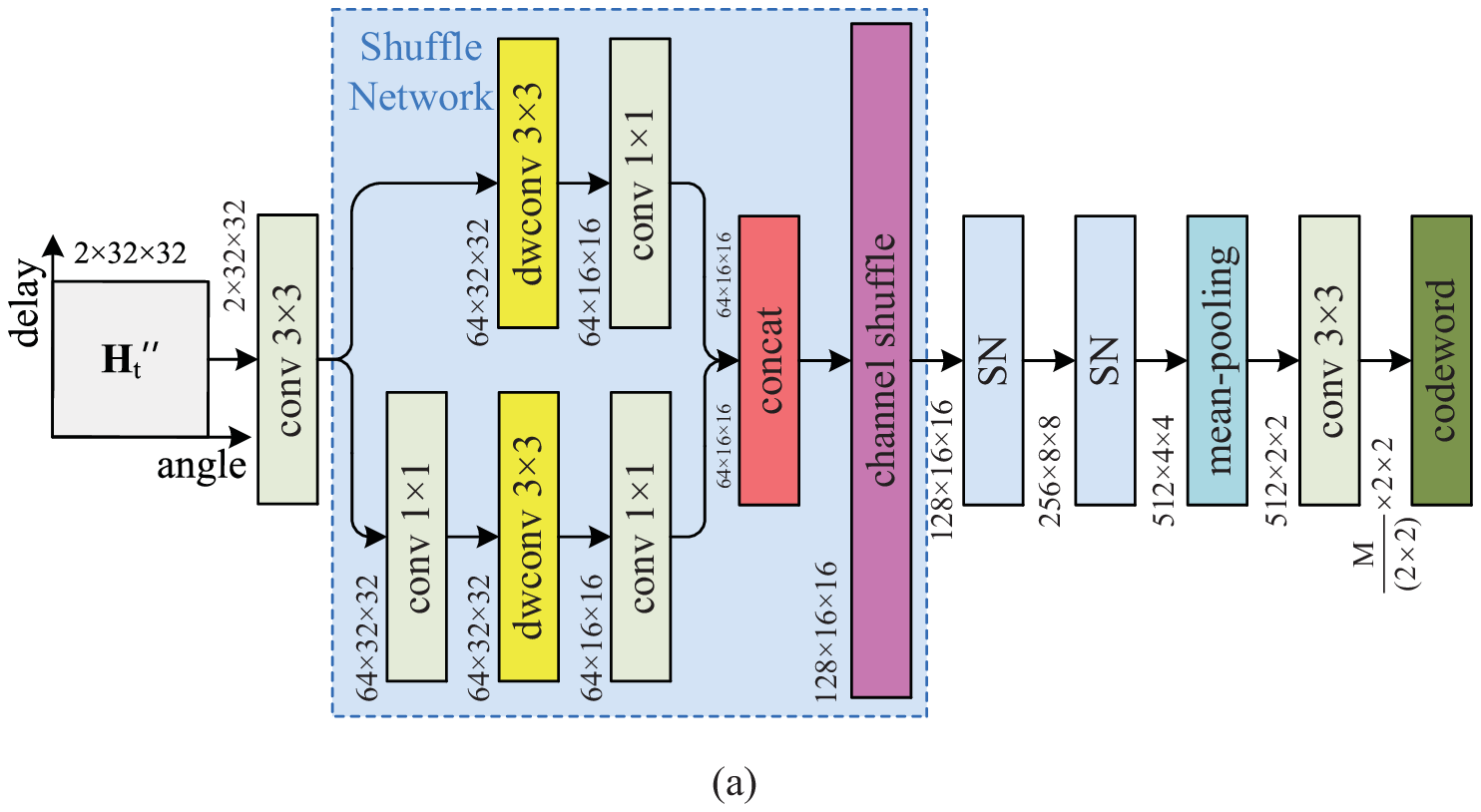}\includegraphics[width=2.619in,height=2.227in]{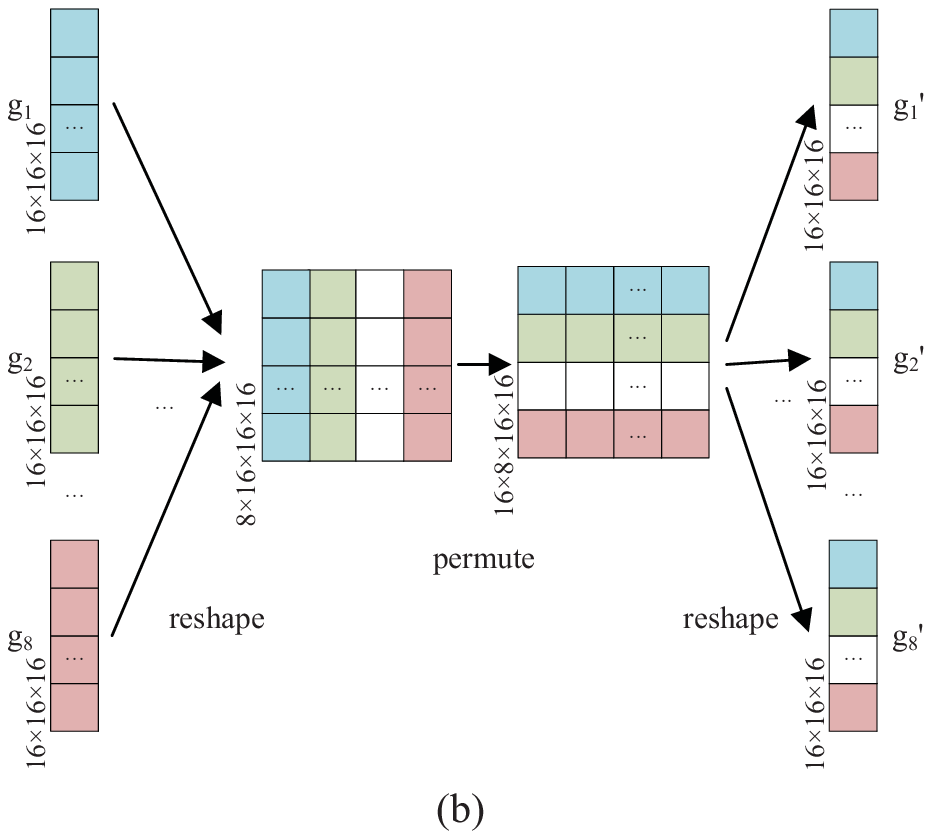}
		\vspace{-0.45cm}
		\caption{(a) Encoder structure of ShuffleCsiNet constructed with convolutional and mean-pooling layers and SN units, which include convolutional, depthwise convolutional, concat, and channel shuffle layers. (b) Schematic of the channel shuffle layers ({\bf conv}: convolutional layer; {\bf dwconv}: depthwise convolutional layer)}
		\vspace{-0.5cm}
		\label{ShuffleCsiNet}
	\end{figure*}

	The CsiNet in \cite{wen2018deep} demonstrates superior performance to traditional CS algorithms considering CSI sensing and recovery. However, this algorithm uses the fully connected layer to change the dimensions. Moreover, CsiNet employs a fully connected autoencoder structure and a fully connected layer to generate codewords. Thus, this network cannot effectively retain the characteristics of 2D image signals. We propose a CSI feedback network based on a convolutional autoencoder called ConvCsiNet to address this issue. The neural network of ConvCsiNet uses convolutional layers instead of fully connected ones to extract features. In theory, the network trained under this architecture can satisfactorily extract the features of 2D image signals. Thus, ConvCsiNet is conducive to improving the reconstruction performance.

	The architecture of ConvCsiNet shown in Fig. \ref{ConvCsiNet} is divided into the following two modules: encoder and decoder. The encoder includes a convolutional layer and four ECN units. The first convolutional layer contains 64 convolutional kernels with a size of $3\times3$ and a sliding step size of 1. The padding methods of this layer is set to "same" to generate a 64-channel feature map with the same size as the input CSI matrix to complete the initial feature extraction. The ECN is used to complete the dimensionality reduction and further feature extraction. Each ECN unit comprises average pooling and convolutional layers. The average pooling layer maintains the number of feature map channels and completes the downsampling which doubles the length and width of the feature map. The convolutional layers in the ECN units use the same features as those of the first convolutional layer. The convolutional layers of all other ECN units, except those of the last unit, maintain the feature map length and width, and the number of channels is doubled to further extract features. The number of channels of the convolutional layer in the last ECN unit is adjusted in accordance with the specific compression ratio. Each convolutional layer is added to the batch normalization (BN) layer and uses the Leaky ReLU activation function. The M-dimensional codeword ${\bf s}$ outputted by the encoder is presented in the form of a feature map with a size of $2\times2$ and the number of channels is M/($2\times2$).

	The decoder part comprises four deconvolution network (DN) units, two RefineNet units, and a convolutional layer. The DN unit is used to restore codeword ${\bf s}$ to the original channel dimension and complete the preliminary reconstruction. Each DN unit comprises a bilinear up-sampling layer and a convolutional layer. The upsampling layer maintains the same number of feature map channels and completes the upsampling which doubles the length and width of the feature map. The convolutional layers in the DN units use a $3\times3$ convolutional kernel with a sliding step of 1 and the padding methods is set to "same". The convolutional layer of the first DN unit increases the number of feature map channels to 512 layers. By contrast, those of the second and third DN units maintain the feature map length and width, and the number of channels is doubled. The convolutional layer of the last DN unit generates a two-channel feature map, that is, two real number matrices, as the initial estimates of the real and imaginary parts of the original CSI matrix. These estimates then enter two similar RefineNet units to further improve the reconstruction quality. The last convolutional layer uses a $3\times3$ convolutional kernel with a sliding step size of 1 and the same padding and a Sigmoid activation function to normalize the output elements to the [0,1] interval. The remaining convolutional layers are added to the BN layer, and the Leaky ReLU activation function is used.

	\subsection{CSI feedback neural network architecture based on lightweight structure}
	\setlength{\parskip}{0.1em}
	The ConvCsiNet aims to optimize the CSI feedback network theoretically despite its complexity. Today, mobile terminals and embedded devices are widely used in the field of modern communications. The parameters and calculations of the ConvCsiNet network model are excessively large. Large memory space and core computing power are necessary to achieve an accurate downlink CSI. Therefore, the lightweight CSI feedback network model is important to ensure a sufficiently accurate training effect. The convolutional layer in ConvCsiNet has numerous parameters and computational complexity. Therefore, we propose a new autoencoder-based CSI feedback network called ShuffleCsiNet. This network replaces the average pooling and the convolutional layers with a lightweight downsampling structure.


    The encoder and decoder parts of the CSI feedback network are located at the UE and BS, respectively. The mobile end has higher parameter requirements than those of the BS. Therefore, ShuffleCsiNet focuses on optimizing the structure of the encoder part, and the decoder part is the same as that in ConvCsiNet.

    Fig. \ref{ShuffleCsiNet}(a) shows that the encoder part of ShuffleCsiNet includes a convolutional layer, three shuffle network (SN) units, and a pooling structure. The first convolutional layer contains 64 convolutional kernels with a size of $3\times3$ and a sliding step size of 1. The padding methods of this layer is set to "same" to generate a 64-channel feature map with the same size as the input CSI matrix to complete the initial feature extraction. The SN is used to complete the dimensionality reduction and further feature extraction. Each SN unit divides the input into the following two parts: one part passes through a $3\times3$ depthwise convolutional layer and a $1\times1$ convolutional layer; the other part successively passes through a $1\times1$ convolutional layer, a $3\times3$ depthwise convolutional layer, and another $1\times1$ convolutional layer. The sliding step of the depthwise convolutional layer is two. The depthwise and the next convolutional layers with respective sizes of $3\times3$ and $1\times1$ can form a depthwise separable convolutional layer. This formed layer uses a $1\times1$ convolution kernel for the linear combination on the channel correlation and then performs convolution on the obtained feature map, which can jointly map the correlation of all dimensions. Moreover, this structure can achieve almost the same effect as the traditional convolution structure but with considerably reduced parameters. Afterward, the two branches pass the concat layer, which halves the size of the feature space and doubles the number of channels to achieve feature reuse. Finally, channel shuffle can ensure information exchange between the two branches.

    The SN unit can double the number of channels while maintaining the length and width of the feature map to further extract features. The pooling structure, which can also complete the downsampling function, comprises average pooling convolutional layers with a size of $3\times3$. Each depth separable convolutional layer is added to the BN layer, and each convolutional layer is added to the BN layer and uses the Leaky ReLU activation function. The M-dimensional codeword ${\bf s}$ outputted by the encoder is presented in the form of a feature map with a size of $2\times2$, and the number of channels is M/($2\times2$).


    The structure of the channel shuffle is shown in Fig. \ref{ShuffleCsiNet}(b). The channel shuffle can reduce element-level operations and increase the information flow after its placement in the concat layer \cite{zhang2018shufflenet}. The most important parameter in the channel shuffle structure is the feature map, that is, the number of shuffle groups. A wide feature map can obtain improved performance benefits within the range allowed by the model. After several experiments, we set the number of shuffle groups to $8$ to obtain satisfactory reconstruction performance.

    Let the transformation formula and all parameters of the entire network be $f(\cdot)$ and $\Theta=\{\Theta_{\sf en},\Theta_{\sf de}\}$, respectively, where the parameters include encoder and decoder parameters. The CSI matrix recovered from the CSI feedback network model proposed in this study can then be expressed follows:

    \vspace{-0.1cm}
	\begin{equation}
	    \begin{aligned}
		\hat{\bf H}''&=f({\bf H}'';\Theta)\triangleq f_{\sf de}(f_{\sf en}({\bf H}'';\Theta_{\sf en}) ;\Theta_{\sf de}).
		\label{CSI matrix}
		\end{aligned}
	\end{equation}

    We use the adaptive moment estimation (ADAM) algorithm to update the parameter set of the network. This algorithm is different from the traditional gradient descent algorithm, which uses a fixed learning rate. The ADAM algorithm can adaptively update the learning rate through training. The loss function of the network is represented by the mean squared error (MSE). Therefore, the prediction loss of the model is defined as follows:

	\vspace{-0.1cm}
	\begin{equation}
		\begin{aligned}
			L(\Theta)= \frac{1}{M}\sum_{m=1}^M\sum_{t=1}^T\|f({\bf H}'';\Theta)-{\bf H}''\|_{2}^{2}.
			\label{loss function}
		\end{aligned}
	\end{equation}
	where $M$ is the total number of samples in the training set and $\| \cdot \|_{2}$ is the Euclidean norm.

\vspace{-0.3cm}
	\section{Numerical results and analysis}
	We compare and analyze the experimental data of ConvCsiNet and ShuffleCsiNet, including their reconstruction performance, parameter amount and FLOPs.

	We use the COST 2100 model \cite{liu2012cost} to obtain the training and testing data and generate channel matrices for two environments. The two environments include indoor cellular and outdoor rural environments in the 5.3 GHz and 300 MHz bands, respectively. We set the bandwidth of the MIMO system to 20 MHz, and the number of subcarriers is $N_{c}$ = 256. A uniform linear array with $N_{t}$ = 32 antennas is used at the BS. The experiments in this study will be performed with CSI compression ratios (CR) of 1/16 and 1/32. The training, validation, and test sets used for offline training contain 75,000, 12,500, and 12,500 samples, respectively.

	The entire training process is performed in the Keras framework. We utilize the ADAM optimizer using MSE loss to configure the training model. The parameters in ADAM are set as $\beta_{1} = 0.9$, $\beta_{2} = 0.999$, and $\epsilon=1 \times 10^{-8}$. This study also uses a small batch training scheme with a batch size of 200, a training round set of 1000, and a learning rate of 0.001. CsiNet, ConvCsiNet, and ShuffleCsiNet are trained and tested on NVIDIA Tesla V100.

	\subsection{CSI reconstruction performance of the networks}
	The normalized MSE (NMSE) is used to evaluate the reconstruction performance and can be defined as follows:


	\vspace{-0.1cm}
	\begin{equation}
		\begin{aligned}
			{\rm NMSE}=\mathbb{E}\bigg \{\frac{1}{T}\sum_{t=1}^{T} {\|{\bf H}''-\hat {\bf H}''\|_{2}^{2}}/{\|{\bf H}''\|_{2}^{2}}\bigg \}.
			\label{nmse}
		\end{aligned}
	\end{equation}
	\vspace{-0.1cm}

	The following cosine similarity is also calculated to facilitate comparison with the CsiNet.

	\vspace{-0.1cm}
	\begin{equation}
		\begin{aligned}
			\rho = \mathbb{E}\Bigg \{\frac{1}{T}\frac{1}{N_{c}} \sum_{t=1}^{T}\sum_{n=1}^{N_{c}}\frac{|\hat {\bf h}_{n}^{H}{\bf h}_{n}|}{\|\hat {\bf h}_{n}\|_{2}\|{\bf h}_{n}\|_{2}} \Bigg \},
			\label{corr}
		\end{aligned}
	\end{equation}
	where $\hat {\bf h}_{n}$ denotes the reconstructed channel vector of the $n$th subcarrier at time $t$, and $\rho$ measures the quality of the beamforming vector when the vector is set as ${\bf v}_{n}={\hat {\bf h}_{n}/\|\hat {\bf h}_{n}\|_{2}}$ because the UE will achieve the equivalent channel ${\hat {\bf h}_{n}^{H}}{\hat {\bf h}_{n}/\|\hat {\bf h}_{n}\|_{2}}$.

		\begin{table}[H]
		\caption{NMSE performance of CSI feedback network designs.}
		\label{performance}
		\centering
		\vspace{-0.1cm}
		\begin{tabular}{m{0.5cm}<{\centering} m{2.2cm}<{\centering} | m{0.75cm}<{\centering} m{0.75cm}<{\centering} m{0.75cm}<{\centering} m{0.75cm}<{\centering}}
			\hline
			\hline
			\multirow{2}{*}{CR} & \multirow{2}{*}{Method} & \multicolumn{2}{c}{Indoor} & \multicolumn{2}{c}{Outdoor} \\
                                &                         & NMSE          & $\rho$      & NMSE        & $\rho$         \\
			\hline
			\multirow{5}{*}{1/16} & CsiNet              & -8.65           & 0.93          & -4.51          & 0.79\\
		  	                    & CRNet               & -11.35          & 0.95          & -5.44          & 0.80\\
			                    & DS-NLCsiNet         & -12.45          & 0.97          & -5.28          & 0.82\\
			                    & \textbf{ConvCsiNet} & \textbf{-13.79} & \textbf{0.98} & \textbf{-6.00} & \textbf{0.85}\\
			                    & ShuffleCsiNet       & -12.14          & 0.97          & -5.00          & 0.82\\
			\hline
			\multirow{5}{*}{1/32} & CsiNet              & -6.24           & 0.89          & -2.81          & 0.67\\
			                    & CRNet               & -8.93           & 0.94          & -3.51          & 0.71\\
		                       	& DS-NLCsiNet         & -8.21           & 0.92          & -3.34          & 0.71\\
			                    & \textbf{ConvCsiNet} & \textbf{-10.10} & \textbf{0.95} & \textbf{-5.21} & \textbf{0.82}\\
			                    & ShuffleCsiNet       & -9.41           & 0.94          & -3.50          & 0.74\\
			\hline
			\hline
		\end{tabular}
	\end{table}
	\vspace{-0.3cm}

    Table \ref{performance} summarizes the performance comparison considering NMSE. We compare the two proposed CSI feedback networks with CsiNet, CRNet \cite{lu2019multi}, and DS-NLCsiNet \cite{yu2020dsnlcsinet}. The table shows that the reconstruction performances of ConvCsiNet and ShuffleCsiNet are superior to that of CsiNet. Compared with the two other CSI feedback networks, ConvCsiNet has an advantage when the compression ratio is 1/32. Notably, ShuffleCsiNet can maintain satisfactory reconstruction performance when parameter quantity and algorithm complexity are substantially reduced. Such a phenomenon is due to the functional reuse in the SN structure of the concat layer and due to the channel shuffle structure, thereby increasing the robustness of the network.


	\subsection{Parameter numbers and FLOPs of the networks}
	The ShuffleCsiNet aims to obtain a light network structure and low algorithm complexity. Therefore, the corresponding reconstruction performance of ShuffleCsiNet is slightly lower than that of ConvCsiNet. This study considers the convolutional , depthwise convolutional , BN layer, and fully connected layers when estimating the parameters of the network model. The parameter numbers of the convolution layer can be expressed as follows:

    \begin{equation}
    N_{\text{conv}} = K^{2} \times C_{i} \times C_{o},
    \end{equation}
	where $K$ is the size of the convolution kernel, $C_{i}$ is the number of input channels, and $C_{o}$ is the number of output channels. The parameter numbers of the depthwise convolutional, BN, and fully connected layers can be respectively expressed as follows:

	\begin{equation}\begin{aligned}
    N_{\text{dwconv}} &= K^{2} \times C_{i}, \\
    N_{\text{bn}} &= 2 \times C_{i}, \\
    N_{\text{fc}} &= C_{i} \times C_{o}.
    \end{aligned}\end{equation}

	Conversely, the convolution, depthwise convolutional, pooling, and fully connected layers are considered when estimating the FLOPS of the network model. The FLOPs of the convolution layer can be expressed as follows:

	\begin{equation}
	O_{\text{conv}} = H_{o} \times W_{o} \times K^{2} \times C_{i} \times C_{o},
	\end{equation}
	where $H_{o}$, $W_{o}$, $C_{i}$, $C_{o}$, and $K$ denote the height of the output, width of the output, number of input channels, number of output channels, and size of the convolution kernel, respectively. The FLOPs of the depthwise convolutional, pooling, and fully connected layers can be respectively expressed as follows:

	\begin{equation}\begin{aligned}
    O_{\text{dwconv}} &= H_{o} \times W_{o} \times K^{2} \times C_{i}, \\
    O_{\text{pool}} &= C_{i} \times H_{i} \times W_{i}, \\
    O_{\text{fc}} &= 2 \times C_{i} \times C_{o}.
    \end{aligned}\end{equation}

	Table \ref{number&FLOPs} shows the model parameter numbers and FLOPs of the two proposed CSI feedback network designs. The model parameters of ShuffleCsiNet are much lower than those of ConvCsiNet. When the CR is 1/32, the parameter amount of ShuffleCsiNet can be reduced to 21.05\% of that of ConvCsiNet. FLOPs can be used to measure the algorithmic complexity of network models. The results show that the complexity of the ShuffleCsiNet algorithm is considerably much lower than that of ConvCsiNet.



	\begin{table}[H]
		\caption{Parameter number and FLOPs of CSI feedback network designs.}
		\label{number&FLOPs}
		\centering
		\vspace{-0.1cm}
		\begin{tabular}{m{0.5cm}<{\centering} m{2.2cm}<{\centering} | m{1.5cm}<{\centering} m{1.5cm}<{\centering}}
			\hline
			\hline
			CR & Method & numbers & FLOPs\\
			\hline
			\multirow{2}{*}{1/16} & ConvCsiNet          & 1,697,144          & 58,515,456\\
			& \textbf{ShuffleCsiNet}       & \textbf{415,528}          & \textbf{11,845,632}\\
			\hline
			\multirow{2}{*}{1/32} & ConvCsiNet          & 1,623,416          & 58,220,544\\
			& \textbf{ShuffleCsiNet}       & \textbf{341,800}          & \textbf{11,550,720}\\
			\hline
			\hline
		\end{tabular}
	\end{table}
	\vspace{-0.3cm}

\vspace{-0.3cm}
	\section{Conclusion}
	\label{conclusion}
	We developed CsiNet and proposed a CSI feedback network called ConvCsiNet in this study based on a convolutional autoencoder. It can effectively extract the features of 2D image signals and exhibits satisfactory reconstruction performance. Subsequently, we proposed a lightweight structured CSI feedback network called ShuffleCsiNet based on ConvCsiNet. The experiments show that the proposed ConvCsiNet displays satisfactory reconstruction performance and the ShuffleCsiNet can substantially save memory space and kernel computing power while ensuring satisfactory reconstruction performance. Both CSI feedback architectures exhibit potentials for practical deployment on realistic MIMO systems.
	\bibliographystyle{IEEEtran}


\end{document}